\documentclass[journal,twocolumn,10pt]{IEEEtran}
\usepackage{graphicx}
\usepackage{verbatim}
\usepackage{upgreek}
\usepackage{amssymb,amsmath}
\usepackage{color}
\usepackage{epstopdf}
\usepackage{bm}
\usepackage{cite}
\usepackage{array,color}
\usepackage{algorithm}
\usepackage{algpseudocode}
\usepackage{amsmath}
\usepackage{amsfonts}
\usepackage{amsthm}
\usepackage{graphics}
\usepackage{epsfig}
\usepackage{soul}
\usepackage{booktabs}
\usepackage{multirow}
\soulregister\cite7
\soulregister\ref7
\usepackage{subfigure}
\usepackage{tabularx}
\usepackage{makecell}

\usepackage{etoolbox}
\makeatletter
\patchcmd{\@makecaption}
{\scshape}
{}
{}
{}
\makeatletter
\patchcmd{\@makecaption}
{\\}
{.\ }
{}
{}
\makeatother

\ifCLASSINFOpdf
\else
\fi

\begin{document}

\title{\huge Joint Beamforming and Offloading Design for Integrated Sensing, Communication and Computation System}

\author{Peng Liu, Zesong Fei,~\IEEEmembership{Senior~Member,~IEEE}, Xinyi Wang,~\IEEEmembership{Member,~IEEE},\\Yiqing Zhou,~\IEEEmembership{Senior~Member,~IEEE}, Yan Zhang,~\IEEEmembership{Member,~IEEE}, Fan Liu,~\IEEEmembership{Senior~Member,~IEEE}
	
	\thanks{Peng Liu, Zesong Fei Xinyi Wang, and Yan Zhang are with the School of Information and Electronics, Beijing Institute of Technology, Beijing 100081, China. (E-mail:bit\_peng\_liu@163.com, feizesong@bit.edu.cn, bit\_wangxy@163.com, zhangy@bit.edu.cn).}
	\thanks{Yiqing Zhou is with the State Key Laboratory of Processors and the Beijing Key Laboratory of Mobile Computing and Pervasive Device, Institute of Computing Technology, Chinese Academy of Sciences, Beijing 100190, China. (E-mail: zhouyiqing@ict.ac.cn)}
	\thanks{F. Liu is with the School of System Design and Intelligent Manufacturing, Southern University of Science and Technology, Shenzhen 518055, China. (E-mail: liuf6@sustech.edu.cn)}
	
}

\maketitle
\begin{abstract}
\textcolor{black}{Mobile edge computing (MEC) is powerful to alleviate the heavy computing tasks in integrated sensing and communication (ISAC) systems.} In this paper, we investigate joint beamforming and offloading design in a three-tier integrated sensing, communication and computation (ISCC) framework comprising one cloud server, multiple mobile edge servers, and multiple terminals. While executing sensing tasks, the user terminals can optionally offload sensing data to either MEC server or cloud servers. To minimize the execution latency, we jointly optimize the transmit beamforming matrices and offloading decision variables under the constraint of sensing performance. An alternating optimization algorithm based on multidimensional fractional programming is proposed to tackle the non-convex  problem. Simulation results demonstrates the superiority of the proposed mechanism in terms of convergence and task execution latency reduction, compared with the state-of-the-art two-tier ISCC framework.
\end{abstract}


\section{Introduction}
\label{sec:intro}
The demands of various emerging services in sixth-generation (6G) wireless networks, such as unmanned aerial vehicle (UAV), vehicular networks, and extended reality, necessitate wireless networks to promptly provide high accuracy sensing and high-throughput communication capabilities \cite{Liu2022jsac}. By supporting simultaneous data transmission and wireless sensing on the same spectrum, the integrated sensing and communication (ISAC) technique is envisioned to effectively enhance hardware and spectral efficiency, and thereby has garnered significant attention from researchers. In \cite{Liu2018TWC}, the employment of separated and shared antenna array were investigated for the ISAC system, where the authors showed that shared antenna array provides better performance in terms of the sensing and communication trade-off. \textcolor{black}{It was then shown in \cite{TSP2020} that the ISAC beamforming is able to provide similar angle estimation performance compared to the optimal sensing-only scheme.} To address the limited communication performance due to the sensing constraint, the authors introduced reconfigurable intelligent surfaces (RIS) into the ISAC systems in \cite{wang2021} to enhance the degrees of freedom and reduce the multi-user interference.

Despite the numerous applications and advantages of the ISAC technique, it also faces challenges in terms of computational complexity. \textcolor{black}{To be specific, the wireless sensing process generates a huge amount of echo signals, resulting in non-negligible  processing delay due to the limited computing capability of ISAC terminals. On the other hand, the sensing tasks are generally delay sensitive, which calls for more efficient sensing task execution framework.} Although reinforcement learning and some other deep learning algorithms have been applied to improve the computing efficiency in \cite{computing1}\cite{computing2}, the training process typically requires substantial computational resources and energy consumption, posing a significant burden on terminal devices. Instead, the mobile edge computing (MEC) technique \cite{MEC}, by allowing terminals to offload sensing tasks to the edge of the network (e.g., base station), is able to alleviate the computational burden of terminal equipments. Following this spirit, the authors in \cite{UAVMEC} investigated the combination of UAV-assisted ISAC and MEC, where UAVs offload collected data to MEC servers while sensing the target. However, the work only considered a single-antenna system, failing to fully utilize spatial degrees of freedom. \textcolor{black}{In \cite{ISCC1}, the authors put forward a general framework of integrated sensing, communication and computation (ISCC) for 6G wireless networks. where the  over-the-air federate learning technique was employed to enhance the computation efficiency.} The authors in \cite{ISCC2} developed an integrated communication, radar sensing and MEC architecture, where user terminals simultaneously execute sensing and computation offloading using multiple-input and multiple-output (MIMO) arrays. \textcolor{black}{However, the aforementioned works only considered  a two-tier computing network including only one MEC server, and have not made full use of multiple  potential MEC servers and cloud servers with abundant computing resources in mobile networks.}

In this paper, we propose a novel three-tier ISCC framework comprised of one cloud server, multiple edge servers, and multiple ISAC user terminals (UTs). Besides locally processing the sensing data, UTs can also offload sensing tasks to either edge servers or the cloud server to  reduce the latency for sensing task processing. We jointly optimize the transmit beamforming matrices at UTs and the offloading decision variables under power budget and sensing signal-to-interference-plus-noise ratio (SINR) constraints. To solve the non-convex problem, we propose an alternating optimization algorithm based on multidimensional fractional programming (MFP) and successive convex approximation (SCA) technique. \textcolor{black}{Numeriacl results validate that the proposed scheme can effectively reduce task execution latency.}

\begin{figure} [t]
	\centering
	\includegraphics[width=3.0in]{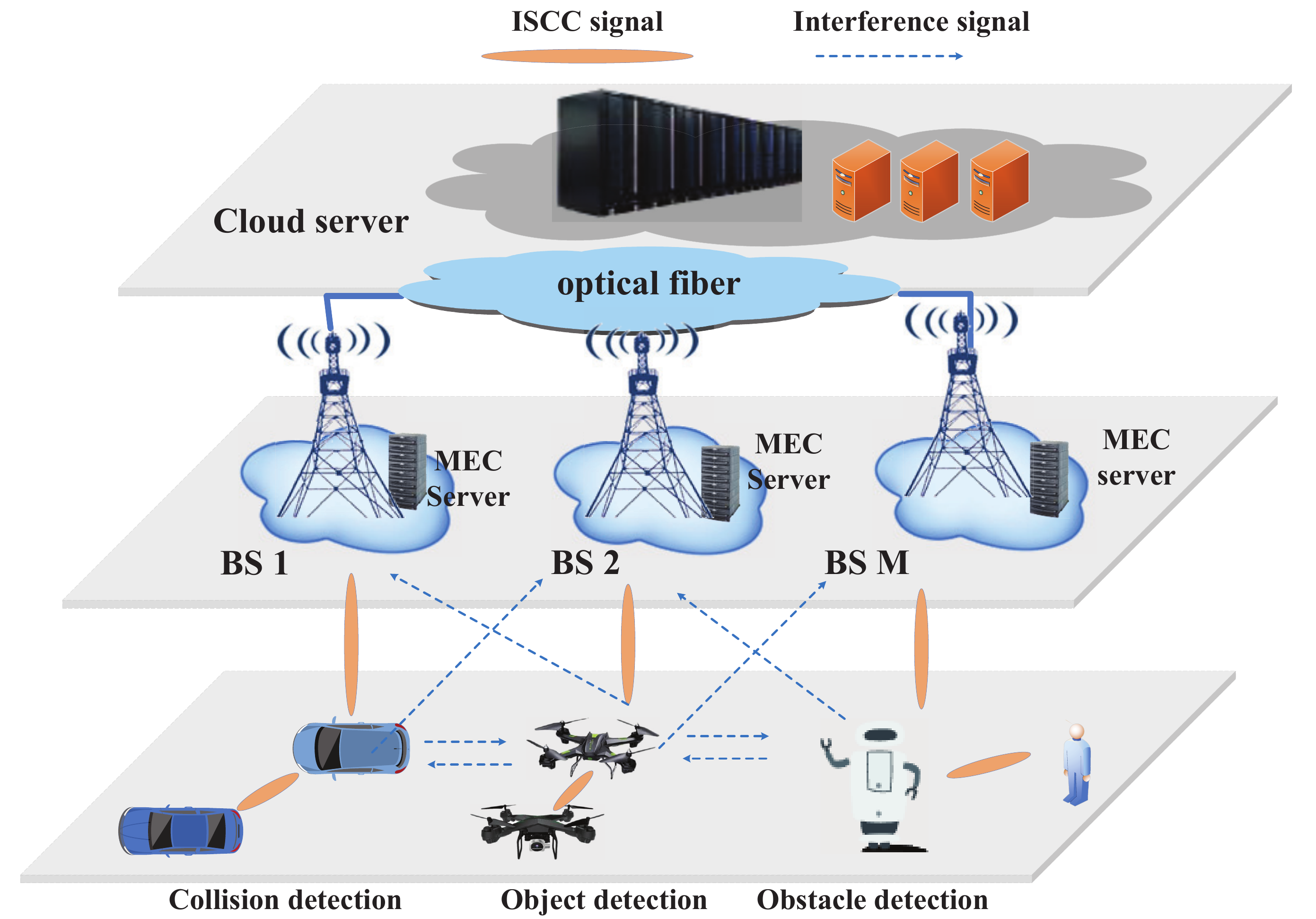}
	\caption{An illustration of the three-tier ISCC system.}
	\vspace{-0.5cm}
	\label{system_model}
\end{figure}

\section{System Model}
\label{sec:format}
We consider a three-tier ISCC system, as depicted in Fig.~1, comprising a cloud server, $M$ base stations (BSs) each equipped with $N$ transmit antennas and $N_r$ receive antennas, and $L$ UTs equipped with $K$ antennas. \textcolor{black}{Here we assume $N_t=N_r=N$ and all UTs operate on the same frequency band.} \textcolor{black}{The BSs are assumed to be able to provide edge computing capabilities for UTs and optionally forward the sensing tasks to cloud through optical fibers.} The UTs could be UAVs, vehicles, or robots engaged in diverse sensing tasks such as target detection/estimation and collision avoidance.  By employing the ISAC technique, UTs can simultaneously execute sensing and task offloading. \textcolor{black}{In particular, we consider the ISAC signals are transmitted via the shared antenna array structure in \cite{Liu2018TWC}.} The received signal at BS $m$ can be expressed as
\begin{equation}
\label{yl} 
\mathbf{y}_{m}=\mathbf{H}_{m,i}^{H}\mathbf{w}_{i}s_{i}+\sum_{l=1,l\neq i}^L\mathbf{H}_{m,l}^{H}\mathbf{w}_{l}s_{l} +\mathbf{z}_m
\end{equation}
where $s_{i}$ denotes the data stream from UT $i$, $\mathbf{w}_{i}\in\mathbb{C}^{K\times d}$ denotes the transmit beamforming matrix of UT $i$, $\mathbf{H}_{m,i}=\in\mathbb{C}^{K \times N}$ denotes the rayleigh channel matrix from  UT $i$ to BS $m$,  and $\mathbf{z}_{m}\sim \mathcal{CN}(0, {\sigma_c}^2\mathbf{I}_N)$ denotes the comlex Gaussian noise at BS $m$.
The data rate from UT $i$ to BS $m$ is given by
\begin{equation}
\label{r1}
R_{m,i} = B\log\det(\mathbf{I}_N+\mathbf{H}_{m,i}^{H}\mathbf{w}_{i}\mathbf{w}_{i}^H\mathbf{H}_{m,i}\mathbf{N}^{-1}_{m,i} ),
\end{equation}
where $B$ is the signal bandwidth. $\mathbf{N}_{m,i}$ is comprised of the covariance matrices of interference from other UTs as well as noise, which is given by
\begin{equation}
\label{r5}
\mathbf{N}_{m,i}={\sum_{l=1,l\neq i}^L\mathbf{H}_{m,l}^{H}\mathbf{w}_{l}\mathbf{w}^H_{l}\mathbf{H}_{m,l}}+{\sigma_c}^2\mathbf{I}_N,
\end{equation}

For the sensing process, the received signals including the echo signal from the target of interest and interference from other UTs is written as
\begin{equation}
\mathbf{r}_i=\alpha_i\mathbf{a}^*({\theta}_i) \mathbf{a}^H({\theta}_i) \mathbf{w}_{i}s_{i}+\sum_{l=1, l \neq i}^L \widehat{\mathbf{H}}_{l,i} \mathbf{w}_{l}s_{l}+\hat{\mathbf{z}}_i,
\end{equation}
where $\widehat{\mathbf{H}}_{l,i}$ represents the interference channel from UT $l$ to UT $i$, $\hat{\mathbf{z}}_i\sim \mathcal{CN}(0, {\sigma_r}^2\mathbf{I}_K)$ is the noise vector at UT $i$,  and $\alpha_i= \sqrt{{\rho}/{{d^r_{i}}^{2}} \times {\zeta}_i/{{d^r_{i}}^{2}}}$ with $\rho$ being the path loss at the reference distance of $d_0=1$m, ${\zeta}_i$ being the radar cross-section (RCS) of UT $i$'s sensing target, and $d^r_{i}$ being the distance from UT $l$ to the sensing target. $\mathbf{a}({\theta}_i)= [1, {\rm e}^{2\pi\Delta\sin(\theta_i)/\lambda}, \cdots, {\rm e}^{2\pi (N -1)\Delta\sin(\theta_i)/\lambda}]^T \in \mathbb{C}^{K \times 1}$ is the steering vector with $\Delta$ being the antenna spacing, ${\theta}_i$ being the angle of arrival corresponding to UT $i$'s target, and $\lambda$ being the wavelength. By defining $\mathbf{A}({\theta}_i)=\mathbf{a}^*({\theta}_i ) \mathbf{a}^H({\theta}_i)$, UT $i$'s echo SINR can be expressed as
\begin{equation}
\label{prob} 
{\rm SINR}^r_i =\frac{\alpha^2\text{tr}(\mathbf{A}({\theta}_i)\mathbf{w}_{i}\mathbf{w}^H_{i}\mathbf{A}({\theta}_i)^H)} {\text{tr}({\sum_{l=1,l\neq i}^L\widehat{\mathbf{H}}_{l,i}^{H}\mathbf{w}_{l}\mathbf{w}^H_{l}\widehat{\mathbf{H}}_{l,i}})+\sigma_r^2},
\end{equation}

\textcolor{black}{Upon receiving the echo signals, UTs have to decide the computation mode for sensing signal processing.} In this paper, three computation strategies are considered, namely local execution, MEC server execution and cloud server execution. We define the sensing data volume as $D_i$  (bit/s), which is linearly correlated with sampling frequency \cite{ISCC2}. Since the sampling frequency is correlated with signal bandwidth, here we assume there exists a linear relationship between data volume and bandwidth. \textcolor{black}{It is worth noting that the size of results is usually smaller than that of the original sensing
data; thus, the time required for computation result feedback is negligible \cite{UAVMEC} \cite{ISCC2}}. The execution time for local computation is expessed as
\begin{equation}  \label{t1}
T_i^L=\frac{\beta{D}_i}{f^L_i}, ~\forall i\in \{1,\cdots,L\},
\end{equation}
where $\beta$ (cycles/bit) denotes the central processing unit (CPU) cycles required for processing one bit of data, which depends on the computational complexity of sensing task, and $f^L_i$ is UT $i$'s CPU frequency  (CPU cycles/s). Under the local execution mode, UT $i$'s power consumption can be expressed as
\begin{equation}
p_i^L=|||\mathbf{w}_{i}||_2+\epsilon(f^L_i)^3,
\end{equation}
\textcolor{black}{where $\epsilon$ is the cofficient determined by the chip architecture \cite{MEC}.} For MEC server execution mode, the total execution time $T_i^C$ for offloading task $D_i$ to the MEC server $m$ is given by
\begin{equation}\small \label{t2}
T_i^M=t_i^\text{M}+t_i^\text{U}=\frac{\beta{D}_i}{f^M_i}+\frac{{D}_i}{R_{mi}},
\end{equation} 
where $t_i^\text{M}$ and $t_i^\text{U}$ represents the computation delay of sensing tasks on the MEC server and the uplink transmission delay from UT $i$ to MEC server $m$, respectively. $f^M_i$ is the CPU frequency allocated to UT $i$ by MEC server $m$. 

Similarly, the total execution time for cloud server computation mode is
\begin{equation} \label{t3}
T_i^C=t_i^\text{C}+t_i^\text{U}+t_i^\text{B}=\frac{\beta{D}_i}{f^C_i}+\frac{{D}_i}{R_{m,i}}+\frac{{D}_i}{r_{c}},
\end{equation}
where ${f^C_i}$ denotes the computation resource allocated to terminal $k$ by cloud server, and $r_c$ is the transmission rate between MEC server and cloud server. Under MEC server execution and cloud server execution modes, UT's  power consumption are both given by
\begin{equation}
p_i^M=p_i^C=||\mathbf{w}_{i}||_2,
\end{equation}

\textcolor{black}{Since the sensing tasks are generally highly integrated and cannot be partitioned, we adopt a binary offloading strategy, where the sensing tasks are executed as a whole either locally at the terminal or offloaded to the MEC/cloud server.} We define the binary offloading decision variables $\bm{a}_m=\{a_{m1},a_{m2}\cdots,a_{mL}\}$ and $\bm{b}_m=\{b_{m1},b_{m2}\cdots,b_{mL}\}$. When $a_{mi}=1$, UT $i$'s sensing task data is offloaded to MEC server $m$, and when $b_{mi}=1$, the sensing task data is offloaded to cloud server by MEC server $m$. We assume each UTs can only select one cmputation mode, i.e.,
\begin{equation} 
\sum_{m=1}^Ma_{mi}+\sum_{m=1}^Mb_{mi}\leq 1, {a_{mi}},{b_{mi}} \in \left\{ {0,1} \right\},~\forall i\in \{1,\cdots,L\}.
\end{equation}

The UT $i$'s total task execution time and power consumption can be given by
\begin{equation}\small
\label{p1}
T_{i}=(1-\sum_{m=1}^Ma_{mi}-\sum_{m=1}^Mb_{mi})T_{i}^L+~\sum_{m=1}^Ma_{mi}T_{i}^M + ~\sum_{m=1}^Mb_{mi}T_{i}^C,
\end{equation}
\begin{equation}\small
\label{p2}
P_{i}=(1-\sum_{m=1}^Ma_{mi}-\sum_{m=1}^Mb_{mi})p_{i}^L+~\sum_{m=1}^Ma_{mi}p_{i}^M + ~\sum_{m=1}^Mb_{mi}p_{i}^C.
\end{equation}

\section{Problem Formulation}
\label{sec:typestyle}
In this section, we aim to jointly design transmit beamforming matrices $\mathbf{w}_i$'s and binary offloading decision variables $\bm{a}_m$'s, $\bm{b}_m$'s to minimize the total execution time while satisfying the sensing SINR and power budget constraints. Specifically, the problem can be formulated as 
{\small
\begin{align}\label{prob29}
\min\limits _{\bm{a}_m,\bm{b}_m, \mathbf{w}_i} & ~\sum_{i=1}^{L}~T_{i}\\
s.t.~~~~& {a_{mi}},{b_{mi}} \in \left\{ {0,1} \right\},~\forall i,m,\tag{\ref{prob29}a},\\
&\sum_{m=1}^Ma_{mi}+\sum_{m=1}^Mb_{mi}\leq 1, ~\forall i, \tag{\ref{prob29}b}\\
& \sum_{i=1}^La_{mi} {f}^M_i \leq {C}_m , \forall m, \tag{\ref{prob29}c}\\
& P_{i} \leqslant P_c,  ~\forall i, \tag{\ref{prob29}d}\\
&{\rm SINR}^r_i\geq \Gamma_{th},\forall i,\tag{\ref{prob29}e}
\end{align}}
where  $(\ref{prob29}\text{c})$ denotes MEC sever's computational capacity  constraint, $C_m$ is the maximum computation resource allocated by the MEC server $m$ to UTs, $P_c$ denotes the maximum power available for each UT, and constraint $(\ref{prob29}\text{e})$ sets a lower bound $\Gamma_{th}$ for the sensing SINR of all UTs.

\section{Algorithm Design}
We observe that it is difficult to solve $(\ref{prob29})$ because of the non-convex objective function, constraints $(\ref{prob29}\text{a})$ and $(\ref{prob29}\text{e})$. Moreover, the coupled variables  $\mathbf{w}_i$ and  $\bm{a}_m, \bm{b}_m$ make the problem even more challenging. Here we propose an alternating optimization algorithm based on MFP and SCA techniques to solve it.

\subsection{Beamforming Matrix Optimization}

With any given $\bm{a}_m, \bm{b}_m$, the objective function of  $(\ref{prob29})$ is determined by $R_{m,i}$. By defining the set of $(m,i)$ offloading pairs as $\mathcal{A}$, the optimization problem can be rewritten as

{\small
\begin{align} 
\label{prob30}
\min \limits _{\mathbf{w}_i} & \sum_{(m,i)\in\mathcal{A}}~\frac{{D}_i}{R_{m,i}}\\
s.t.~
&(\ref{prob29}\text{d}),(\ref{prob29}\text{e}), \notag
\end{align}}
which is a multiple-ratio FP problem. To make problem (\ref{prob30}) more tractable, by introducing auxiliary variables $c_{i}$'s, we rewrite the problem (\ref{prob30}) as
{\small
	\begin{align} 
\label{prob31}
\min \limits _{\mathbf{w}_i,c_i} & \sum_{(m,i)\in\mathcal{A}}~c_{i}\\
s.t.~
&{R_{m,l}}\geq D_i/c_{i},~~\forall (m,i) \in\mathcal{A},\tag{\ref{prob31}a}\\
&(\ref{prob29}\text{d}),(\ref{prob29}\text{e}). \notag
\end{align}}

\textcolor{black}{From (\ref{r1}), we observe that the optimization variables $\mathbf{w}_i$ within $R_{mi}$ are located inside the function $\log\det(\cdot)$, rendering it non-convex.} We begin by transforming $R_{mi}$ into a concave form. $R_{m,i}$ can be expressed as \cite{MEC} 
\begin{equation}
\label{r2}
\widehat{R}_{m,i} = B\log(1+\mathbf{w}^{H}_{i}\mathbf{H}_{m,i}\mathbf{N}^{-1}_{m,i}\mathbf{H}_{m,i}^{H}\mathbf{w}_{i} ).
\end{equation}

By applying the quadratic transform \cite{fp4}, (\ref{r2}) can be futher reformulated as
\begin{equation} 
\label{r3}
\tilde{R}_{m,i} =\max_{\bm{z}_{i}} B\log(1+2\Re\{\bm{z}^H_{i}\mathbf{H}^{H}_{m,i}\mathbf{w}_{i}\}-\bm{z}^H_{i}\mathbf{N}_{m,i}\bm{z}_{i}),
\end{equation}
where $\bm{z}_{i}$ are auxiliary variables. Note that the SINR term in (\ref{r2}) has been converted into a concave function with respect to $\mathbf{w}_{k}$ in (\ref{r3}). Therefore, $\bm{z}_{i}$'s and $\mathbf{w}_{i}$'s can be optimized in an iterative way. In each iteration. $\bm{z}_{i}$ is optimally updated by 
\begin{equation}
\label{r4}
\bm{z}_{i} =\mathbf{N}^ {-1}_{m,i}\mathbf{H}_{m,i}\mathbf{w}_{i},
\end{equation}
and substituted into (\ref{r3}). Next, we reformulate the non-convex constraint (\ref{prob29}\text{e}) as
\begin{equation}
\label{prob2} \small 
{\text{tr}(\mathbf{w}^H_{i}\mathbf{A}({\theta}_i)^H}\mathbf{A}({\theta}_i)\mathbf{w}_{i})\geq \frac{\Gamma_{i}}{\alpha^2}{{\sum_{l=1,l\neq i}^L\text{tr}(\mathbf{w}^H_{l}\widehat{\mathbf{H}}_{l,i}\widehat{\mathbf{H}}_{l,i}^{H}\mathbf{w}_{l})}+\sigma_r^2}.
\end{equation}

By emloying the SCA technique at a given point $\widetilde{\mathbf{w}}_i$, the non-convex constraint (\ref{prob2})  can be linearized as
\begin{equation}
\begin{aligned}
\label{prob3} 
{ 2\text{tr}(\widetilde{\mathbf{w}}^H_{i}\mathbf{A}({\theta}_i)^H\mathbf{A}({\theta}_i)\mathbf{w}_{i})-\text{tr}(\widetilde{\mathbf{w}}^H_{i}\mathbf{A}({\theta}_i)^H\mathbf{A}({\theta}_i)\widetilde{\mathbf{w}_{i}})} \\
\geq \Gamma_{th}/\alpha^2{{\sum_{l=1,l\neq i}^L\text{tr}(\mathbf{w}^H_{l}\widehat{\mathbf{H}}_{l,i}\widehat{\mathbf{H}}_{l,i}^{H}\mathbf{w}_{l})}+\sigma_r^2},
\end{aligned}
\end{equation}

Based on the above transformations, the problem (\ref{prob30}) is reformulated as
{\small
\begin{align} 
\label{prob32}
\min\limits _{\mathbf{w}_i,{c}_{i}} & \sum_{(k,l)\in\mathcal{A}}~c_{i}\\
s.t.~
&~{\tilde{R}_{mi}}\geq D_i/c_{i},~~\forall (m,i) \in\mathcal{A},\tag{\ref{prob32}a}\\
&~(\ref{prob29}\text{d}),(\ref{prob29}\text{e}),(\ref{prob3}) \notag
\end{align}}
which can be solved via the interior point method \cite{2004Convex}. By iteratively handling problem (\ref{prob32}) and updating $\bm{z}_{i}$ as (\ref{r4}), the original sum of ratios problem (\ref{prob30}) can be solved. 
\renewcommand{\algorithmicrequire}{\textbf{Input:}}
\renewcommand{\algorithmicensure}{\textbf{Output:}}
\begin{algorithm}[t] 
	\caption{Alternating Optimization Algorithm for Solving (\ref{prob29}) }
	\begin{algorithmic}[1] 
		\Require  Sensing SINR threshold $\Gamma_{th}$, power budget $P_{c}$,
		
		\State {Initialize offloading decision variables $\bm{a}_m, \bm{b}_m$, auxiliary variables $\bm{z}_{i}$ and transmit beamforming matrices $\widetilde{\mathbf{w}}_i$ to a feasible value.}
		\Repeat
		\Repeat
		\State {Update $\mathbf{w}_i^{}$'s by solving problem (\ref{prob32}).}
		\State {Update $\bm{z}_{i}$ by equation (\ref{r4}).}   
		\Until  {the objective value in (\ref{prob32}) converge.}
		\State {Obtain optimized relaxed solution of $\bm{a}_m$ and $\bm{b}_m$ by solving the problem (\ref{prob33}).}
		\State {Recover $\bm{a}_m, \bm{b}_m$ to 0-1 variables.}
		\Until {the objective value in (\ref{prob29}) converge.}
		
		\Ensure the beamforming matrices $\mathbf{w}_i, \forall i$ and offloading decision variables $\bm{a}_m, \bm{b}_m$.
	\end{algorithmic}
\end{algorithm}
\subsection{Offloading Decision Optimization}
To handle the binary variables in constaint $(\ref{prob29}\text{a})$, we relax $\bm{a}_i, \bm{b}_i$ into $0\leq{a_{mi}}\leq 1,0\leq {b_{mi}}\leq 1$. With given $\mathbf{w}_{i}$, the problem (\ref{prob29}) can be expressed as
\begin{align} 
\label{prob33}
\min\limits _{\bm{a}_i,\bm{b}_i} & ~\sum_{i=1}^{L}~T_{i}\\
s.t.~
&~0\leq{a_{mi}}\leq 1,0\leq {b_{mi}}\leq 1,\tag{\ref{prob33}a}\\&~(\ref{prob29}\text{b}),(\ref{prob29}\text{c}),(\ref{prob29}\text{d}), \notag
\end{align}
which is a linear programming problem and can be solved directly using interior point method. After obtain the relaxation solution, we employ the relax continuous and inflation method in \cite{Luong2017} to recover  0-1 variables.

\subsection{Overall Algorithm }
By iteratively solving problem (\ref{prob32}) and (\ref{prob33}), the original problem (\ref{prob29}) can be solved. To make it clearer, we summarize the alternating optimization algorithm for solving (\ref{prob29}) in Algorithm 1. 

Next, we focuse on analyzing the convergence of Algorithm 1. For subproblem (\ref{prob32}), we first define the objective fuction as $\tilde{f}(\mathbf{w},\bm{z})$. Since problem (\ref{prob32}) is a convex problem, the objective value is non-increasing during the optimization process, i.e.,
\begin{equation}
\label{c1}
\tilde{f}^{(iter+1)}(\mathbf{w}^{(iter+1)},\bm{z}^{(iter)})\leq \tilde{f}^{(iter)}(\mathbf{w}^{((iter))},\bm{z}^{(iter)}), 
\end{equation}
where $(iter)$ denotes the index of iteration. 

For the update of $\bm{z}^{(iter)}_{i}$, we first re-express $\tilde{R}_{m,i}$ as $B\log(1+\bm{z}^H_{i}\mathbf{H}_{m,i}\mathbf{w}_{i}+\mathbf{w}^{H}_{i}\mathbf{H}^H_{m,i}\bm{z}_{i}-\bm{z}^H_{m,i}\mathbf{N}_{m,i}\bm{z}_{i})$, which is further rewritten as $B\log(1+\mathbf{w}^{H}_{i}\mathbf{H}_{m,i}^{H}\mathbf{N}^{-1}_{m,i}\mathbf{H}_{m,i}\mathbf{w}_{i}-(\bm{z}^H_{i}-\mathbf{w}^{H}_{i}\mathbf{H}^H_{m,i}\mathbf{N}^ {-1}_{m,i})\mathbf{N}_{m,i}(\bm{z}_{i}-\mathbf{N}^ {-1}_{m,i}\mathbf{H}_{m,i}\mathbf{w}_{i})$. It is easy to observe that by updating $\bm{z}^{(iter)}_{i}$ via (\ref{r4}), the left side of constraint (\ref{prob32}a) is non-decreasing. Thus, we have
\begin{equation}\small
\label{c2}
\tilde{R}^{(iter+1)}_{i}(\mathbf{w}^{(iter+1)}_i,\bm{z}^{(iter+1)}_{i})\geq \tilde{R}^{(iter+1)}_{i}(\mathbf{w}^{(iter+1)}_i,\bm{z}^{(iter)}_{i}).
\end{equation}

It is apparent that the equality within constraint (\ref{prob32}a) holds when the optimal value is achieved for problem (\ref{prob32}). As $\tilde{R}_{m,i}$ increases, $c_i$ decreases. Therefore, we have 
\begin{equation}\small
\label{c3}
\tilde{f}^{(iter+1)}(\mathbf{w}^{(iter+1)}_,\bm{z}^{(iter+1)})\leq \tilde{f}^{(iter)}(\mathbf{w}^{(iter+1)},\bm{z}^{(iter+1)}).
\end{equation}

Since the optimal value in each iteration is non-decreasing and  lower-bounded by a finite value, subproblem (\ref{prob32}) is guaranteed to converge. At the same time, since the objective (\ref{prob33}) is also non-increasing \cite{Luong2017} and the optimal value of the original problem  (\ref{prob29}) is lower-bounded, Algorithm 1 is guaranteed to converge.

The computational complexity of Algorithm 1 comprises three main components involving update $\bm{z}_{i}$ and solving subproblems (\ref{prob32}) and (\ref{prob33}). Refering to \cite{Wang2014}, the overall complexity of Algorithm 1 is $\mathcal{O}\left(N_{iter}(K^{3}L^{3}+M^3L^3+N^3L+KN^2L)\right)$, where $N_{iter}$ represents the number of iterations.

\begin{figure} [t]
	\centering
	\includegraphics[width=2.8in,height=2.1in]{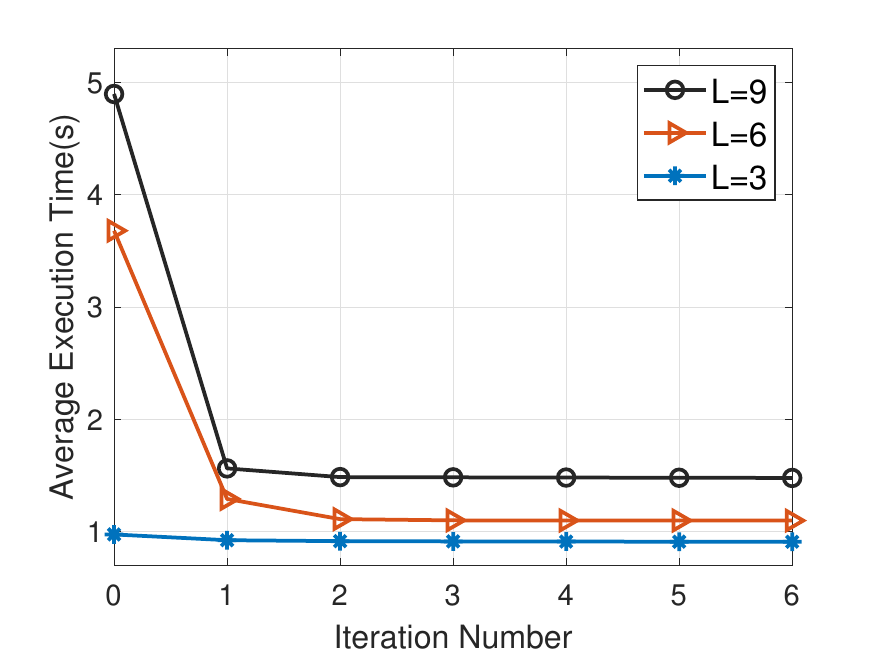}
	\caption{Average Sensing Task Execution Time v.s. Iteration Number.}
	\label{fig2}
\end{figure}

\section{Simulation Results}
In this section, we demonstrate the effectiveness of the proposed three-tier computation offloading scheme integrating beamforming optimization (THCO-BO). We set the antennas numbers of UTs and BSs as $K=12$ and $N=16$, respectively. The computing frequency of UTs, MEC server, and cloud server are set as $f_i^L=0.1$~Gcycles/s, $f_i^M=3$~Gcycles/s, and $f_i^C=10$~Gcycles/s, respectively. $\beta$ is set as 40~cycles/bit \cite{ISCC2}. The numbers of BSs and UTs are $M=3$ and $L=9$, respectively. Three other schemes, including two-tier (MEC and UTs) computation offloading scheme with beamforming optimization (TTCO-BO) \cite{ISCC2}, three-tier computation offloading scheme with maximal ratio communication (THCO-MRC, $\mathbf{w}_i=\sqrt{P_c}\frac{\mathbf{H}_{mi}[\lambda_{max}]}{||\mathbf{H}_{mi}||}$) and  three-tier computation offloading scheme with maximal ratio sensing (THCO-MRS, $\mathbf{w}_i=\sqrt{P_c}\frac{\mathbf{a}({\theta}_i)}{||\mathbf{a}({\theta}_i)||}$) are carried out for comparisons, where  $\lambda_{max}$ represents the column corresponding to the largest eigenvalue of $\mathbf{H}_{mi}$.

Fig.~2 illustrates the convergence performance of the proposed algorithm for different users, demonstrating its fast convergence speed. Additionally, as the UT number $L$ increases, inter-user interference intensifies, resulting in decreased uplink rates and longer average execution time. 

Fig.~3 shows the variation of the average sensing task execution time with different ISAC signal bandwidths. Note that larger bandwidths result in higher uplink transmission rates, but also entail a larger volume of echo signals. We can observe that as the data volume increases, the average execution time of all schemes increases. Nevertheless, when the BS  has sufficient computational resources $C_m$ for UTs, the TTCO-BO and THCO-BO schemes exhibit similar sensing task execution performances, outperforming the local execution scheme. However, when computational resources available at BSs are limited, the proposed THCO-BO scheme outperforms TTCO-BO scheme. This is because the proposed TTCO-BO scheme offers more selectivity in dealing with sensing tasks and can adjust its offloading strategy to different resource budget.

Fig.~4 shows the average execution time under different sensing thresholds $\Gamma_{th}$ and different transmit power budgets. Due to the decrease of inter-user interference, the proposed THCO-BO scheme yields lower task execution latency compared to THCO-MRC and THCO-MRS schemes. As the sensing SINR threshold $\Gamma_{th}$ rises, \textcolor{black}{there is a corresponding increase in average execution time, indicating the trade-off between sensing accuracy and processing delay. Moreover, as the power budget increases, the degrees of freedom in beamforming optimization increase, resulting in the increase of offloading rate and the reduction of execution latency.}

\begin{figure}[t]
	\centering
	\includegraphics[width=2.8in,height=2.1in]{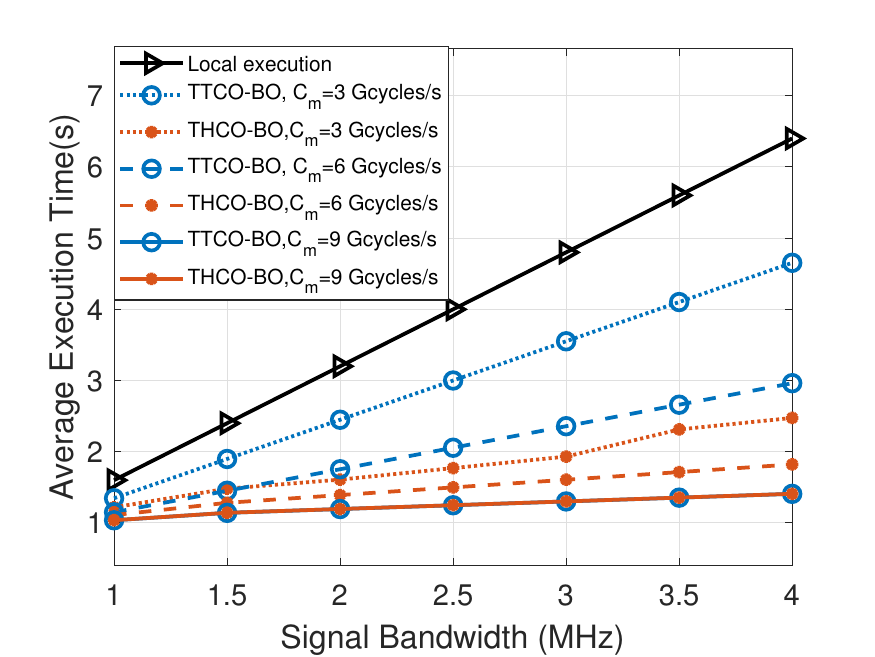}
	\caption{Average Sensing Task Execution Time v.s. ISAC Signal Bandwidth.}
	\label{fig2}
\end{figure}
\begin{figure}[t]
	\centering
	\includegraphics[width=2.8in,height=2.1in]{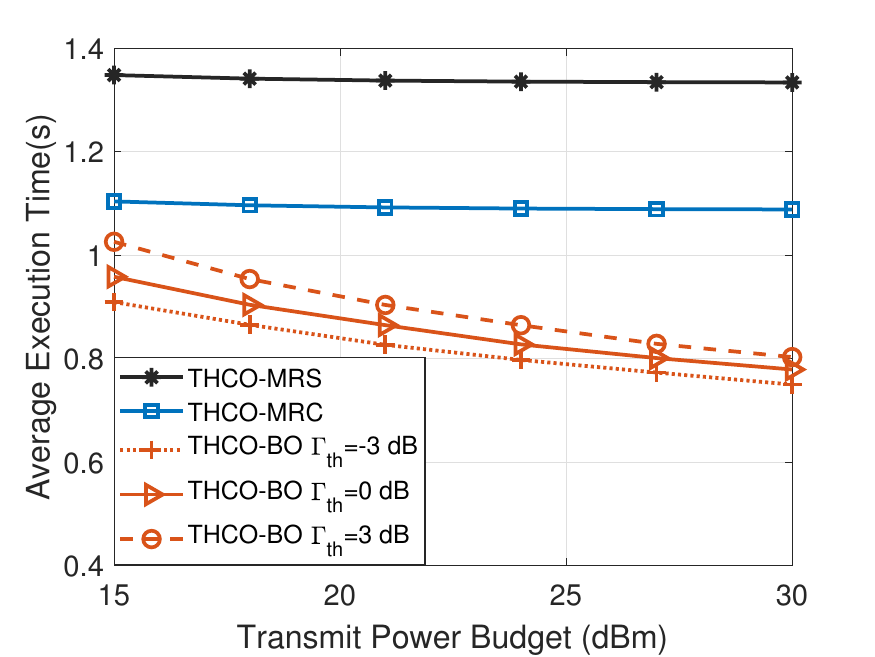}
	\caption{Average Sensing Task Execution Time v.s. Transmit Power Budget.}
	\label{fig2}
\end{figure}

\section{Conclusion}
In this paper, we considered a three-tier ISCC framework comprised of one cloud server, multiple MEC servers, and multiple UTs, where UTs can optionally offload computationally intensive sensing data to either MEC servers or the cloud server while sensing targets. We aimed to minimize the total task execution latency under sensing performance and power budget constraints. Specifically, we jointly optimized the transmit beamforming matrix and offloading decision variables. Through elaborate transformation, the non-convex joint design can be solved by an alternating algorithm based on MFP and SCA techniques. Simulation results demonstrated the effectiveness of the proposed framework in reducing task execution latency compared to existing schemes.


\bibliographystyle{IEEEtran}
\bibliography{biblp/bibfilelp}

\end{document}